\documentclass[10pt,letterpaper]{article}
\usepackage{opex3}

\begin{document}

\title{A novel method for polarization squeezing with Photonic Crystal Fibers}

\author{Josip Milanovi\'c$^{1,2}$, Mikael Lassen$^{1,3}$, Ulrik L. Andersen$^{1,3}$ \\ and Gerd Leuchs$^{1,2}$}

\address{$^1$Max Planck Institute for the Science of Light, G\"{u}nther-Scharowsky-Str. 1, 91058 Erlangen, Germany\\
$^2$Institute of Optics, Information and Photonics, University of Erlangen-Nuremberg, Staudtstr. 7/B2, 91058 Erlangen, Germany\\
$^3$Department of Physics, Technical University of Denmark, 2800 Kongens Lyngby, Denmark}

\email{Josip.Milanovic@mpl.mpg.de} 


\begin{abstract}
Photonic Crystal Fibers can be tailored to increase the effective Kerr nonlinearity, while producing smaller amounts of excess noise compared to standard silicon fibers. Using these features of Photonic Crystal Fibers we create polarization squeezed states with increased purity compared to standard fiber squeezing experiments. Explicit we produce squeezed states in counter propagating pulses along the same fiber axis to achieve near identical dispersion properties. This enables the production of polarization squeezing through interference in a polarization type Sagnac interferometer. We observe Stokes parameter squeezing of $-3.9 \, \pm0.3 \, \mathrm{dB}$ and anti-squeezing of $16.2 \, \pm0.3 \, \mathrm{dB}$.
\end{abstract}

\ocis{(270.0270) Quantum optics, (270.6570) Squeezed states, (270.2500) Fluctuations, relaxations, and noise, (270.5290) Photon statistics, (060.5295) Photonic crystal fibers.} 


\bibliographystyle{osajnl}
\bibliography{OE-PCF}


\section{Introduction}

The generation of squeezed states of light in optical glass fibers, via the optical Kerr effect $(\chi^{(3)})$, has, to date, been greatly influenced by various effects, which either introduce large amounts of excess noise or even hinder efficient creation of squeezed light. The excess noise, which manifests itself primarily as phase noise originates from inelastic scattering of photons on acoustic phonons (Brillouin scattering) and optical phonons (Raman scattering). Especially Guided Acoustic Wave Brillouin Scattering (GAWBS) has been investigated intensively as it is believed to be the dominant effect responsible for phase noise in optical glass fibers at least for lower peak powers \cite{dong_experimental_2008,heersink_efficient_2005,corney_many-body_2006,shelby_broad-band_1986}. Photonic Crystal Fibers (PCFs) \cite{russell_photonic_2003} represent a promising new technology to reduce this noise \cite{elser_reduction_2006} and are therefore excellent candidates for the production of squeezed states with higher purity which is of crucial importance for numerous quantum communication and information protocols \cite{braunstein_quantum_2003}. Not only the reduction of GAWBS and scattering effects are important in order to achieve quantum states of light with higher purity also the spatial, spectral and temporal overlap between different light pulses needs to be considered. In the case of fiber squeezing low interference visibility can result from light pulses that experience different dispersion within the birefringent fibers. Therefore, in the experiment presented here all light pulses propagate along the same fiber axis which makes the quality of the interference less dependent on the dispersion properties of the fiber, see Fig.\ref{ExpSetup} for the experimental realization.

Since the Kerr effect conserves the photon number, the amplitude fluctuations remain at the shot noise level preventing direct detection of the squeezing. However, this problem can be overcome by using a experimental setup such as the nonlinear optical loop mirror (NOLM) \cite{fiorentino_soliton_2002,kitagawa_number-phase_1986,schmitt_photon-number_1998,drummond_quantum_1993} or by performing the quantum noise measurement in observables which are easier to access such as the polarization states of light. When restricting to second order correlations \cite{Sanchez-Soto_Quantum_2000} the quantum polarization states can be described by its three Stokes observables \cite{stokes_composition_1852,jauch_theory_1955,robson_theory_1974,siegman_lasers_1986} which obey the standard commutation relations for angular momenta. Since a coherent state is not an eigenstate of the Stokes observables all polarization measurements are limited by quantum noise. Therefore polarization squeezing is associated with the squeezing of Stokes observables to below the fluctuations of the standard quantum noise limit (QNL) \cite{Korolkova2005}. The theory was first suggested in the work by Chirkin \emph{et al}. in 1993 \cite{chirkin_quantum_1993}, where the Heisenberg uncertainty relations for the Stokes operators were derived, followed by the experimental realization in optical fibers \cite{korolkova_polarization_2002,bowen_experimental_2002,heersink_polarization_2003}. The Poincare sphere \cite{born_principles_1999} provides a convenient way of representing the Stokes parameters as well as their quantum noise. The upper and lower poles represent left and right-circularly polarized light ($\pm S_3$). Points on the equatorial plane indicate linear polarization. Note that if the only non vanishing average Stokes parameter is $S_3$ (circular polarization) the 'polar region' can be projected onto the equatorial plane spanned by the $S_1$ and $S_2$ parameters as longs as their uncertainty is not too high. The squeezing and anti-squeezing of the quantum polarization can be investigated in this plane. In this limit, there is a clear analogy between polarization squeezing and quadrature squeezing \cite{josse_polarization_2003}.

A squeezed polarization state is an element of the Hilbert state spanned by two orthogonally polarized but otherwise identical spatio-temporal modes. The measurement of the Stokes parameters involves projections within this two-mode space \cite{dong_experimental_2008,heersink_efficient_2005,sorensen_quantum_1998,lassen_generation_2007,josse_polarization_2003}. When attempting to create a squeezed polarization state the two initially separate spatio-temporal modes have to interfere. Low interference contrast introduces vacuum and excess noise, respectively, reducing the amount of detectable squeezing in the quantum state. The interference contrast of these two modes is a crucial parameter for the efficiency of polarization squeezing. 
In the present paper we report on the generation of polarization squeezing with PCFs using an improved method. We employed pulses counter propagating along the same fiber axis to achieve identical dispersion properties. This improved method also simplifies the experimental setup, since no compensation is needed for the pulse delay caused by the birefringence of the polarization maintaining fiber \cite{heersink_polarization_2003,silberhorn_generation_2001}. The squeezing generated with this setup less dependent on the dispersion properties of the fiber.

\section{Experimental procedure}

\begin{figure}[tbp]
\begin{center}
\includegraphics[width=0.65\textwidth]{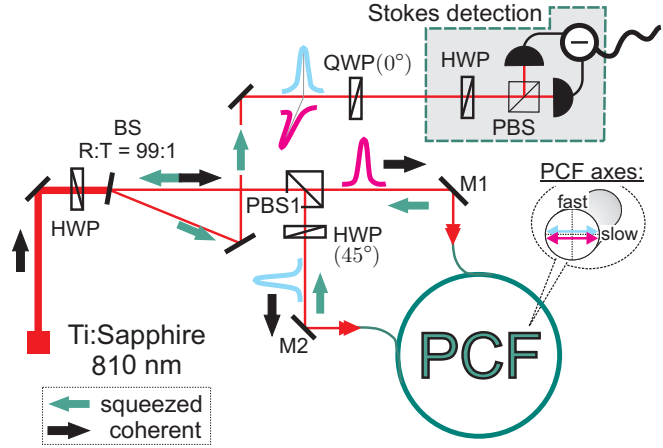}
\caption{Schematic of the experimental squeezing setup. BS: beam splitter PBS: polarizing beam splitter. HWP: half-wave plate. QWP: quarter-wave plate.}
\label{ExpSetup}
\end{center}
\end{figure}

The schematics of the experimental setup is depicted in Fig.\ref{ExpSetup}. We use 1 meter of single mode polarization maintaining NL-PM-750 PCF (Crystal Fibre A\slash S) with a zero dispersion wavelength at $750 \, \mathrm{nm}$. The PCF uses a micro-structured cladding region with air holes to guide light in the pure silica core. The PCF supports a mode with an effective mode field diameter of $1.8 \, \pm0.2 \, \mu\mathrm{m}$, which yields an enhanced effective nonlinearity due to the strong light localization. Ultra short laser pulses with approximately $120 \,\mathrm{fs}$ pulse duration are used to exploit the Kerr nonlinearity $(\chi^{(3)})$ of the fiber. The pulses are generated with a commercial 'Tsunami' Ti:Sapphire laser from Spectra Physics Inc. at a central wavelength of $810 \, \mathrm{nm}$. The pulse repetition rate is 82~MHz and the average output power is approximately $2\, \mathrm{W}$.

A linearly polarized beam is equally divided and coupled into both ends of the PCF which is placed in a polarization type Sagnac interferometer. For the in- and out-coupling of the optical field we use aspheric lenses with numerical apertures of 0.41 and a focal length f = 4.5~mm. The half-wave plates (HWPs) are aligned such that the polarization of the two counter propagating beams are aligned along the same axis of the polarization maintaining PCF. This was checked by monitoring the dark port of PBS1. 
This is an essential point of the setup in order to assure that both pulses have the same output spectra. After recombination of the two beams with a relative phase yielding linearly polarized light, they are directed to the Stokes detection scheme via a 99:1 beam-splitter (BS). Here the two pulses pass through a quarter wave-plate (QWP) which turns the linear polarization into a circular polarization. The squeezing and anti-squeezing is then simply measured in the 'dark' {$S_1$-$S_2$} plane. The {$S_1$ and $S_2$} Stokes parameters can be accessed by rotating the HWP and subtracting the resulting photo currents of the two outputs of the PBS. This measurement is equivalent to a balanced homodyne detection, where the bright excitation acts as the local oscillator for the orthogonally polarized dark mode. The spectral densities of the resulting difference currents are measured with an electronic spectrum analyzer (ESA).

\subsection{Visibility and spectral evolution}

\begin{figure}[h]
\begin{center}
\begin{tabular}[b]{lr}
\includegraphics[width=0.45\textwidth,keepaspectratio]{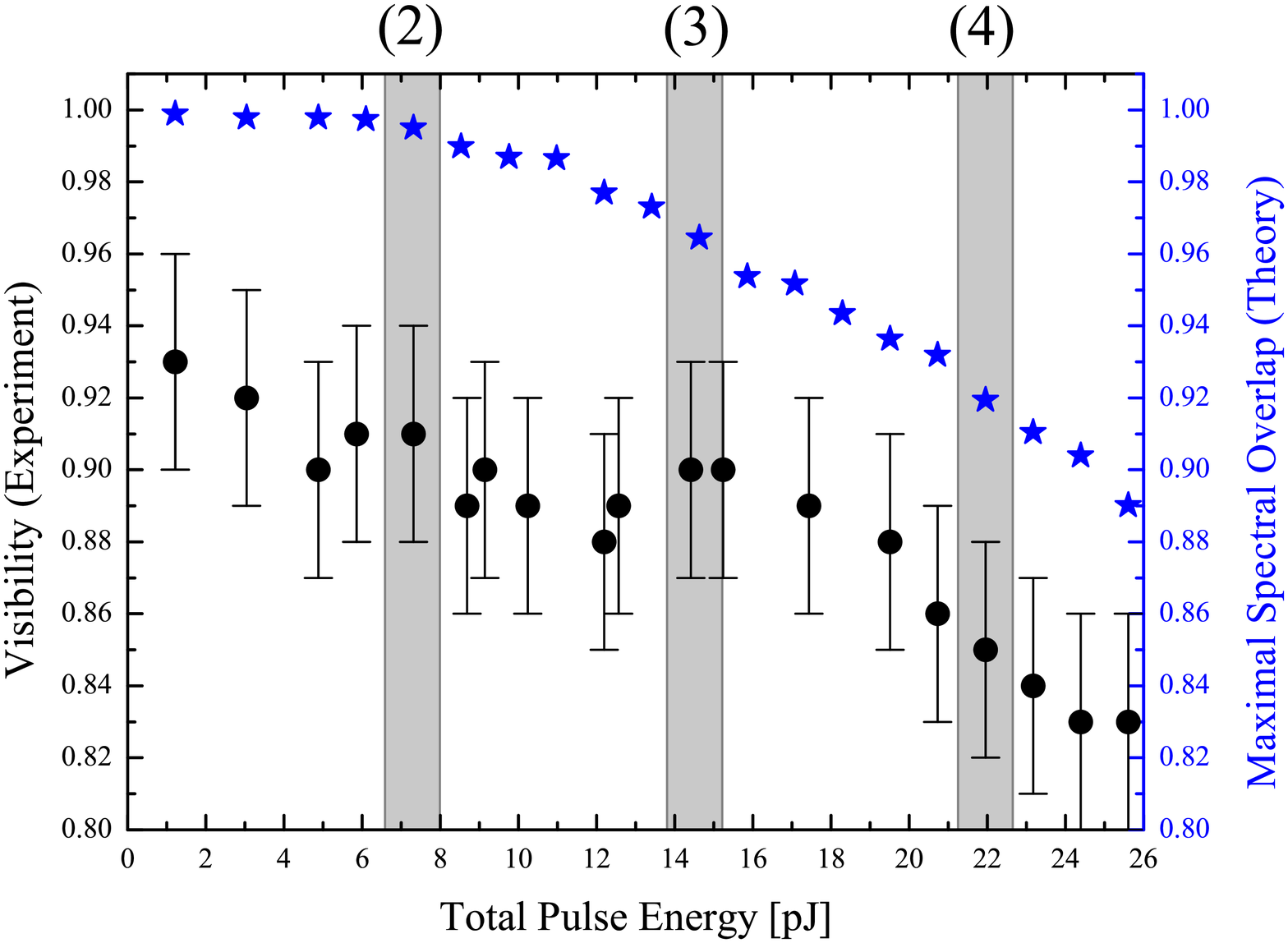} & \includegraphics[width=0.40\textwidth,keepaspectratio]{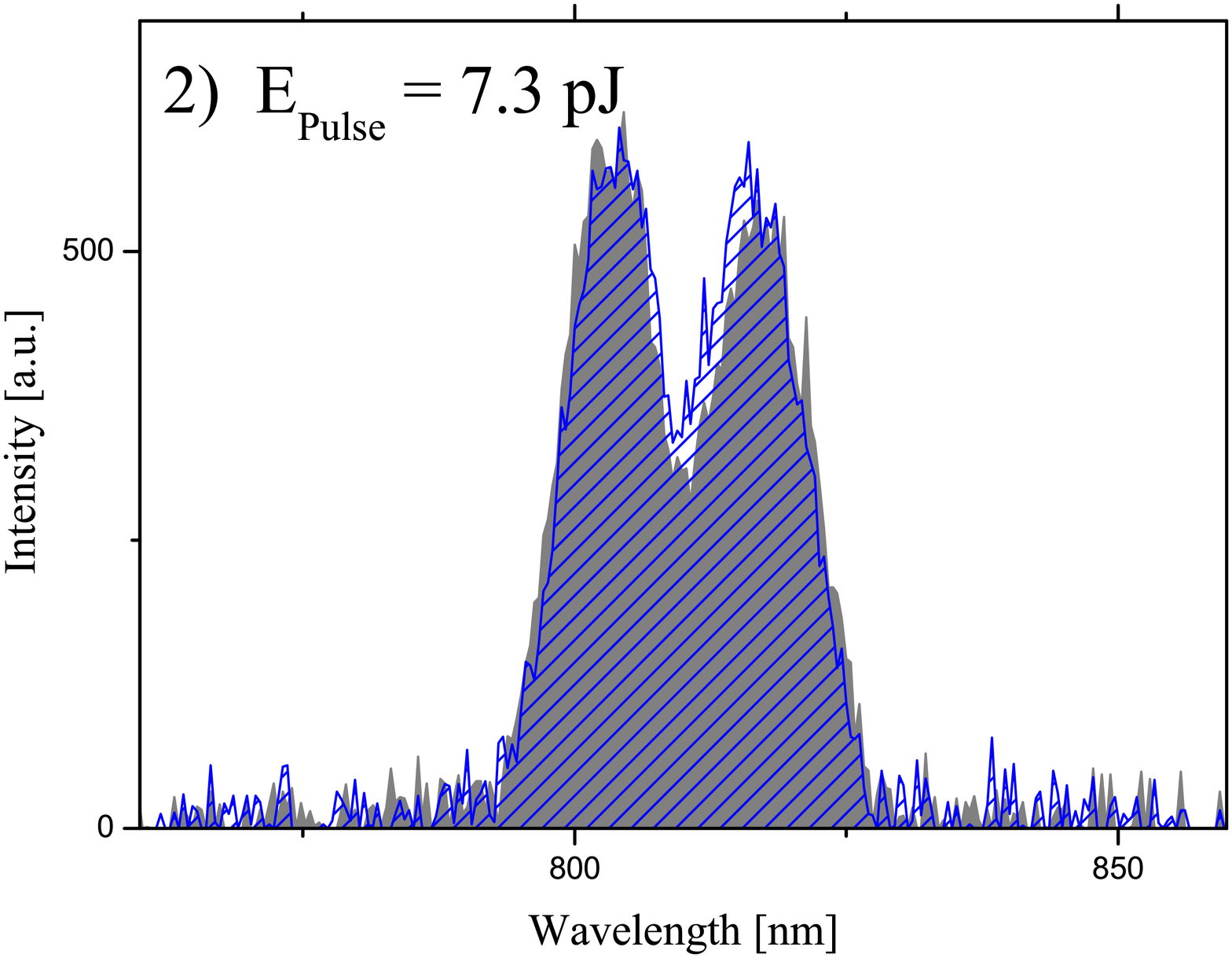} \\\\
\includegraphics[width=0.40\textwidth,keepaspectratio]{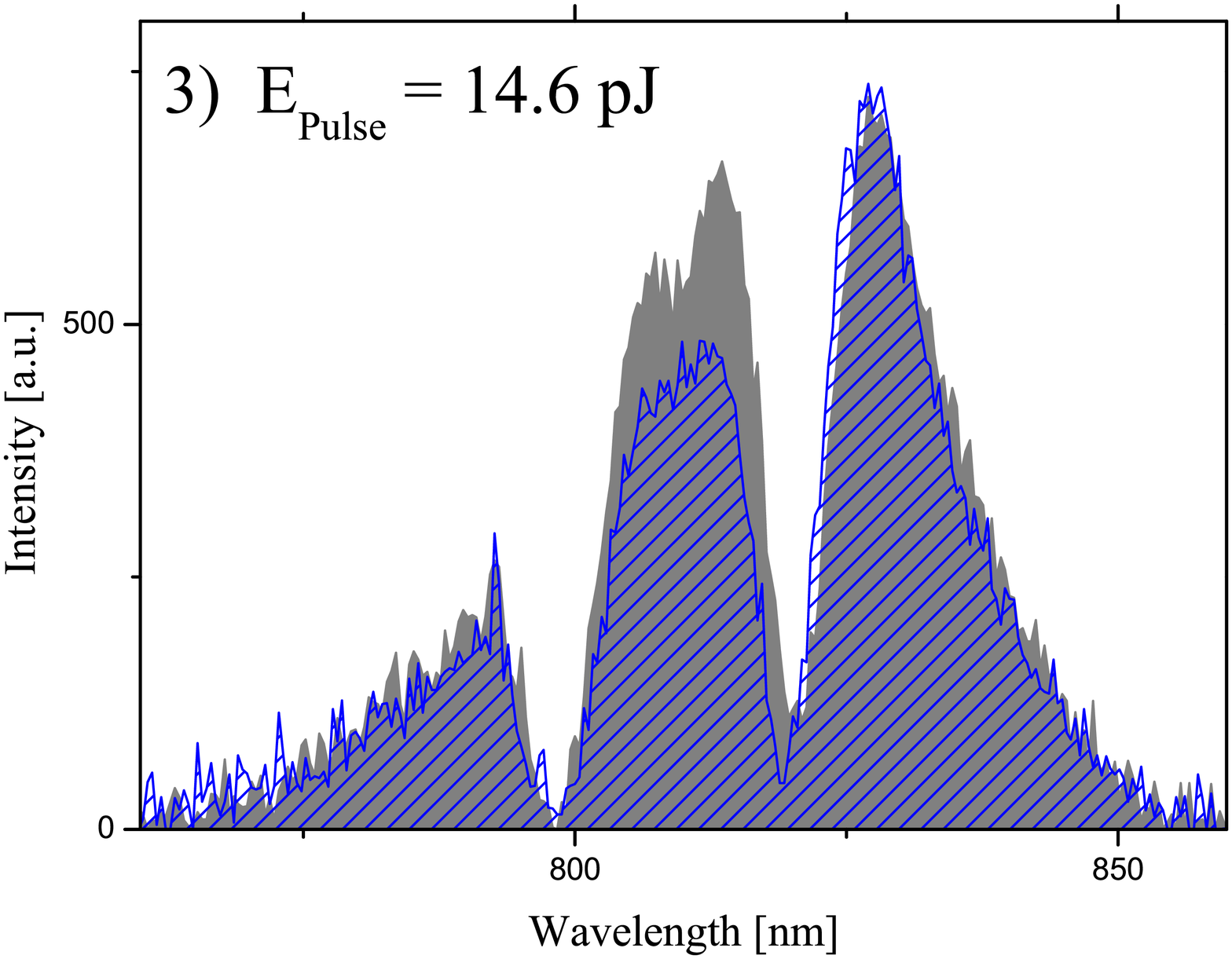} & \includegraphics[width=0.40\textwidth,keepaspectratio]{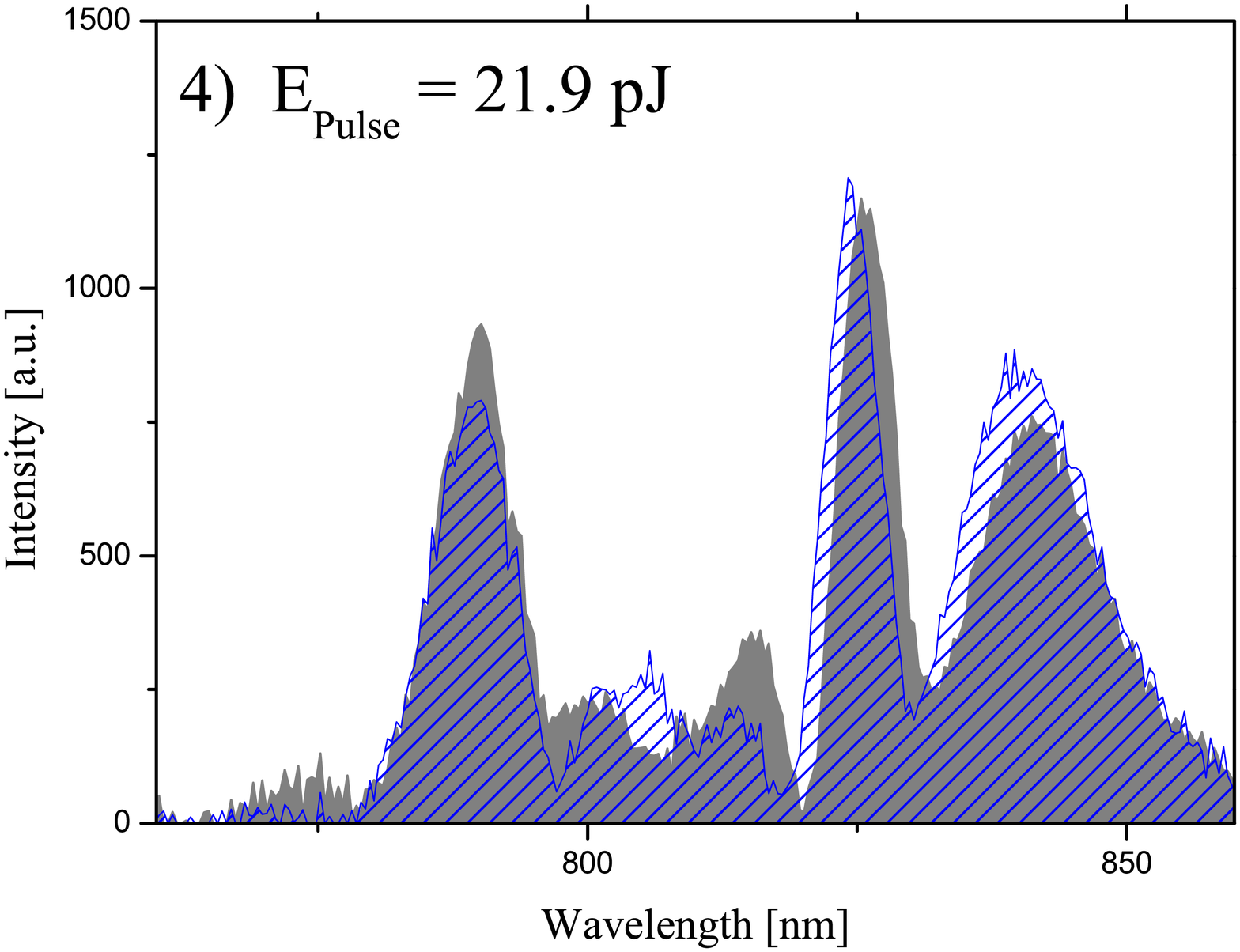}
\end{tabular}
\caption{The black circles show the experimentally measured visibility (left scale) whereas the blue stars show the theoretical maximum for the spectral overlap $\mathcal{V}_{\mathrm{max}}$ of the corresponding spectra (right scale). The marked areas (2)-(4) are pulse energies for which the measured output spectra are shown in detail: (2)\, $7.3 \, \mathrm{pJ}$, (3)\, $14.6 \, \mathrm{pJ}$ and (4)\, $21.9 \, \mathrm{pJ}$. The gray and blue shaded areas are the s- and p-polarized outputs from PBS1, respectively. (NL-PM-750 fiber, $810 \, \mathrm{nm}$ center input wavelength.)}
\label{visibility}
\end{center}
\end{figure}

Since the dispersion parameters of the orthogonal fiber axes are different, the spectral evolution of the orthogonal polarized pulses will also be different. As a consequence the pulses evolving along different fiber axes can not interfere perfectly and the amount of detectable squeezing will decrease with increasing pump power. In our previous work this effect degraded the efficient generation of polarization squeezing \cite{milanovic_polarization_2007}. Here we observe that the spectral evolution of the two counter propagating pulses on the same optical axis is quite similar, as expected, which is due to the identical dispersion response. For maximal interference of the pulses we have to make sure that they have equal power when they leave the fiber. Due to the imperfect and unequal incoupling into the PCF the pulses can have different powers inside the PCF and hence experience different nonlinear evolution. This leads to a different spectral evolution for the counter-propagating pulses. Using an optical spectrometer we measured the spectra of the outcoming beams at each end of the fiber. The results for different powers are shown in Fig.~\ref{visibility}. The visibility was measured by removing the quarter wave plate QWP and measuring the maximum and minimum light power in one of the output ports of the Stokes detection PBS. By manipulating the incoupling mirrors $M_1$ and $M_2$, which also were the outcoupling mirrors, the visibility could be tweaked and measured as shown in Fig.~\ref{visibility}. Note that the inaccuracy of the visibility measurement (4-5\%) mainly stems from fluctuations in the power measurement.

We observe that with increasing pulse energy, spectral broadening occurs. The spectral broadening is due to a complex interaction of several linear and nonlinear effects; e.g. dispersion, self-phase modulation, Brillouin and Raman scattering \cite{corwin_fundamental_2003}. The different spectral evolution of the counter-propagating pulses leads to a decrease in the spectral overlap of the two pulses. For characterizing the visibility we chose the spectral overlap $\mathcal{V}$ \cite{saleh_fundamentals_2007} as figure of merit. The spectral overlap only considers the characteristic of the spectra in contrast to the experimental visibilty which also considers other attributes as difference in phase, spatial mode or temporal mode. For the counter propagating fields ($E_{\mathrm{p}}$, $E_{\mathrm{s}}$), which are orthogonally polarized after the PBS1, we have calculated an upper bound of the spectral overlap $\mathcal{V}_{\mathrm{max}}=\frac{\int\mathrm{d}\omega \, |E_{\mathrm{p}}(\omega)|\,|E_{\mathrm{s}}(\omega)|}{\frac{1}{2} \left( \int\mathrm{d}\omega\, |E_{\mathrm{p}}(\omega)|^2 + \int\mathrm{d}\omega \, |E_{\mathrm{s}}(\omega)|^2 \right)}$. Fig.~\ref{visibility} shows the measured visibility and theoretical maximal spectral overlap $\mathcal{V}_{\mathrm{max}}$ as a function of the pump power. It can be seen that the experimentally measured visibility show the same behavior as the calculated overlap. However, the reason for the experimental values not reaching the same high degree of overlap is that the theoretical values assume no phase difference, perfect temporal and spatial overlap which, in general, is not the case in the experiment.

In the experiment the interference contrast (visibility) of the two Kerr squeezed modes turnes out to be very sensitive. In conventional single-pass fiber squeezing experiments the spatial mode-matching occurs automatically and the temporal mode-matching is normally controlled with an active feedback loop. 

In our experiment the in-coupling into the individual fiber ends are imperfect and not identical, the out-coupling angle can therefore slightly differ for both sides. The reason for the different in- and outcoupling is probably because of imperfect optics and that the pump beams are not identically collimated. This will lead to a noticeable difference in phase, temporal and spatial overlap of the counter propagating pulses. Also small mechanical fluctuations of the PCF ends lead to imperfect mode matching, since the fiber ends fluctuate independently. This lead us to believe that the dominant effect which causes the difference between the experimentally measured visibility and the theoretically predicted spectral overlap is the spatial mismatch of the two modes. As the spatial mismatch is power independent it is consistent with the fact that there is a nearly constant offset between the data from the theory and the experiment (Fig.~\ref{visibility}).

Compared to our previous work \cite{milanovic_polarization_2007}, where the spectral overlap between the corresponding pulses was very low, we have increased the spectral overlap of the pulses with this new experimental procedure. As the detection efficiency scales quadratically with interference visibility, even small increases are an important improvement.

For this experiment glass ferules with approximately 1.5\,mm core diameter have been cleaved to the PCF ends to enhance the in-coupling. We have to exclude the possibility that the ferules corrupt the optical mode. By measuring squeezing and visibility without these ferules we always stayed clearly below the results which were measured using the ferules. This is probably due to mechanical vibrations which effect the in-coupling into the bare PCF more than when using glass ferules. If we consider the lower visibility without the ferrules the extrapolated amount of squeezing is the same as with using the ferrules.

\subsection{Measurement of polarization squeezing}

\begin{figure}[tbp]
\begin{center}
\includegraphics[width=0.7\textwidth]{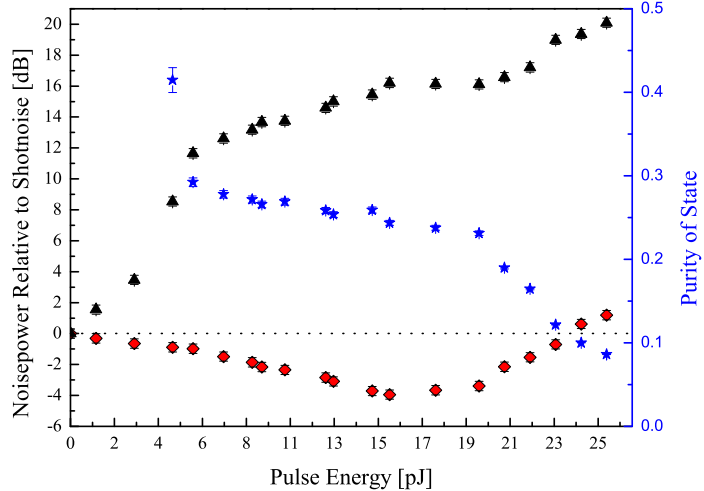}
\caption{Measured noise power versus the optical pulse energy. Red squares and black triangles show squeezing and anti-squeezing, respectively (left scale). We observed a maximal squeezing of $-3.9 \,\pm0.3\,\mathrm{dB}$ with an anti-squeezing of $16.2 \, \pm0.3 \, \mathrm{dB}$. The blue stars show the purity of the squeezed state (right scale). All data is measured at a frequency of 17~MHz. The squeezing is only corrected for electronic detector noise which is $13 \, \mathrm{dB}$ below the shot noise.}
\label{POL SQL}
\end{center}
\end{figure}

The measured squeezing and anti-squeezing versus total pulse energy are plotted in Fig.~\ref{POL SQL}. With increasing
pulse energy the squeezing increases until a certain point after which the squeezing decreases. The measured noise even exceeds the shot noise level for higher pulse energies. This excess noise is mainly composed of GAWBS, Raman scattering as well as uncorrelated modes which were created by nonlinear four-wave-mixing processes. We observed a soliton squeezing of $-3.9 \, \pm0.3 \, \mathrm{dB}$ with an excess noise of $16.2 \, \pm0.3 \, \mathrm{dB}$ (Fig.\ref{POL SQL}). All measurements were performed at a detection frequency of 17 MHz. The variances have all been corrected for dark-noise of the detectors, which is more than $13 \, \mathrm{dB}$ below the QNL $0 \, \mathrm{dB}$. Each trace depicted in Fig.~\ref{POL SQL} is normalized to the QNL. The measurements are performed with a resolution bandwidth of the electronic spectrum analyzer (ESA) set to 300 kHz and with a video bandwidth of 300 Hz. The calibration of the QNL is done by sending a coherent beam with equal power to that of the squeezed beam into the Stokes measurement.

The total detection efficiency of our experiment is given by: $\eta_{\mathrm{total}}=\eta_{\mathrm{prop}}\eta_{\mathrm{det}}\eta_{\mathrm{vis}}$, where
$\eta_{\mathrm{prop}}=0.95\pm0.01$ is the propagation efficiency from the fiber to the detectors, $\eta_{\mathrm{det}}=0.95 \pm0.05$ is the quantum efficiency of the photo-detectors and $\eta_{\mathrm{vis}}$ accounts for the non unity visibility in our Stokes measurements (see Fig.~\ref{visibility}) ranging from $(0.83)^2\pm 0.04$ to $(0.93)^2\pm 0.04$. The purity $\mathcal{P}$ of the squeezed state is calculated as $\mathcal{P}={\left[\Delta^2 \hat{S}{(\mathrm{sqz})} \cdot \Delta^2 \hat{S}{(\mathrm{antisqz})} \right]}^{-1/2}$. By correcting for propagation losses after the fiber and interference losses between the polarization modes, the maximum inferred squeezing is $-8.7 \, \pm0.8 \, \mathrm{dB}$ and the anti-squeezing is $18.5 \, \pm0.8 \, \mathrm{dB}$. Comparing the measured purity (Fig.~\ref{POL SQL}) to the purity of squeezed states generated with standard fibers \cite{dong_experimental_2008}, the purity of our squeezed state is notably higher, approximately 3-4 times, although we could not reach the same amount of squeezing as in the work by R. Dong \emph{et. al.} \cite{dong_experimental_2008}. We attribute this increase in purity to the fact that the light pulses accumulate less phase noise while propagating along the PCF. On one hand the microstructure is known to reduce phase noise \cite{elser_reduction_2006}. On the other hand phonon-photon interactions which scale with fiber length are avoided due to the much shorter fiber. This length reduction is possible as the PCFs have higher effective Kerr nonlinearities compared to standard silicon fibers.

\section{Conclusion}
We have demonstrated the generation of polarization squeezed light using a PCF. We generated $-3.9 \,\pm0.3\,\mathrm{dB}$ squeezing and $16.2 \,\pm0.3\,\mathrm{dB}$ anti-squeezing. We successfully exploited a Sagnac-loop type setup with counter propagating pulses along the same fiber axis.
Compared to our previous work, where the spectral overlap between the corresponding pulses was very low, in this experiment we increased the spectral overlap and also increased the amount of measured squeezing. The interference contrast of the two Kerr squeezed modes is highly sensitive to imperfect optical components and mechanical fluctuations of the PCF ends. The future task is to achieve a constant high visibility to produce a reliable and simple to use source for bright entangled states.

This work was supported by the EU project COMPAS. Mikael Lassen acknowledges support from the Alexander von Humboldt Foundation.
\end{document}